\documentclass[12pt]{article}
\usepackage[dvips]{graphicx,color}
\usepackage{amsmath,bm,calc}
\usepackage[psamsfonts]{amssymb}
\begin{document}

\begin{titlepage}
\title{Application of Gaussian expansion method
to nuclear mean-field calculations with deformation}

\author{H. Nakada\\
\textit{Department of Physics, Graduate School of Science,
 Chiba University,}\\
\textit{Yayoi-cho 1-33, Inage, Chiba 263-8522, Japan}}

\date{\today}
\maketitle
\thispagestyle{empty}

\begin{abstract}
We extensively develop a method
of implementing mean-field calculations for deformed nuclei,
using the Gaussian expansion method (GEM).
This GEM algorithm has the following advantages:
(i) it can efficiently describe the energy-dependent asymptotics
of the wave functions at large $r$,
(ii) it is applicable to various effective interactions
including those with finite ranges,
and (iii) the basis parameters are insensitive to nuclide,
thereby many nuclei in wide mass range can be handled
by a single set of bases.
Superposing the spherical GEM bases
with feasible truncation for the orbital angular momentum
of the single-particle bases,
we obtain deformed single-particle wave-functions
to reasonable precision.
We apply the new algorithm to
the Hartree-Fock and the Hartree-Fock-Bogolyubov calculations
of Mg nuclei with the Gogny interaction,
by which neck structure of a deformed neutron halo
is suggested for $^{40}$Mg.
\end{abstract}


\vspace*{3mm}\noindent
\end{titlepage}

\pagestyle{plain}

\section{Introduction}
\label{sec:intro}

As experimental facilities supply fruitful data
on nuclei far off the $\beta$-stability,
it has been recognized that theoretical approaches to nuclear structure
should be renewed in some respects.
One of the key ingredients is
wave-function asymptotics at large $r$,
which sometimes produces neutron (proton) halos
in vicinity of the neutron (proton) drip line.
It is also important to reinvestigate effective interactions,
in connection to the magic numbers
that are different from those near the $\beta$-stability line.
We developed a method
for the spherical mean-field calculations~\cite{ref:NS02,ref:Nak06}
in which the Gaussian expansion method (GEM)~\cite{ref:Kam88}
was applied.
This method seems suitable to studying the shell structure
from stable to drip-line nuclei,
owing to its several advantages.

A large number of nuclei have quadrupole deformation.
The deformation plays a significant role in unstable nuclei as well.
For instance, the deformation can be relevant
to the disappearance of the $N=20$ magic number
on the so-called `island of inversion'~\cite{ref:IoI},
whereas there remain arguments for individual nuclei;
\textit{e.g.} spherical description has also been proposed
for $^{32}$Mg~\cite{ref:Mg32-sph}.
It is desired to implement calculations with deformation,
reproducing the wave-function asymptotics
and handling a wide variety of effective interactions simultaneously.

In this paper we propose a new method
for deformed mean-field calculations by applying the GEM.
Taking advantage of the flexibility
in describing the radial degrees of freedom,
we adopt a set of the spherical GEM bases
to represent deformed single-particle (s.p.) wave functions,
with truncation for the orbital angular momentum $\ell$.
The new method is tested
in the Hartree-Fock (HF) and the Hartree-Fock-Bogolyubov (HFB)
calculations for Mg nuclei with the Gogny interaction,
and the results are compared with those in literatures.
From the present calculation, an interesting feature is suggested
for the drip-line nucleus $^{40}$Mg.

\section{Single-particle bases}
\label{sec:basis}

In this paper we assume the nuclear mean fields to be axially symmetric
and to conserve the parity.
The $z$ axis is taken to be the symmetry axis.
The method can immediately be extended to general cases
with no symmetry assumptions on the one-body fields,
apart from an additional constraint on the center-of-mass (c.m.) position.

\subsection{GEM bases}
\label{subsec:basis-set}

We represent the s.p. wave functions
by superposing the spherical Gaussian bases,
which have the following form:
\begin{eqnarray} \varphi_{\nu\ell jm}(\mathbf{r})
&=& R_{\nu\ell j}(r)\,[Y^{(\ell)}(\hat{\mathbf{r}})
\chi_\sigma]^{(j)}_m\,; \nonumber\\
R_{\nu\ell j}(r) &=& \mathcal{N}_{\nu\ell j}\,r^\ell\exp(-\nu r^2)\,.
\label{eq:basis} \end{eqnarray}
Here $Y^{(\ell)}(\hat{\mathbf r})$ expresses the spherical harmonics
and $\chi_\sigma$ the spin wave function.
We drop the isospin index without confusion.
The range parameter of the Gaussian basis $\nu$
is a complex number in general~\cite{ref:GEM};
$\nu=\nu_\mathrm{r}+i\nu_\mathrm{i}$ ($\nu_\mathrm{r}>0$).
Via the imaginary part oscillating behavior of the s.p. wave functions
can be expressed efficiently~\cite{ref:Nak06}.
Formulae for calculating the one- and two-body matrix elements
that are required in the HF and the HFB calculations,
as well as the constant $\mathcal{N}_{\nu\ell j}$,
are given in Refs.~\cite{ref:NS02,ref:Nak06}.
The s.p. wave functions under the axially deformed mean field
are represented as
\begin{equation}
 \psi_{n\pi m}(\mathbf{r}) = \sum_{\nu\ell j} c^{(n)}_{\nu\ell jm}\,
  \varphi_{\nu\ell jm}(\mathbf{r})\,,
 \label{eq:spwf}
\end{equation}
where the subscript $\pi$ on the lhs stands for the parity.
The sum of $\ell$ and $j$ on the rhs runs over
all possible values satisfying $\pi=(-)^\ell$, $j=\ell\pm 1/2$
and $j\geq |m|$, in principle.

In the GEM we usually take $\nu$'s
belonging to a geometric progression.
In Ref.~\cite{ref:Nak06}, we found that a certain combination
of the real- and complex-range Gaussian bases
is suitable for nuclear mean-field calculations.
In all the following calculations,
we take the basis-set of
\begin{equation}
\nu_\mathrm{r}=\nu_0\,b^{-2n}\,,\quad
\left\{\begin{array}{ll}\nu_\mathrm{i}=0 & (n=0,1,\cdots,5)\\
{\displaystyle\frac{\nu_\mathrm{i}}{\nu_\mathrm{r}}
=\pm\frac{\pi}{2}} & (n=0,1,2)\end{array}\right.\,,
 \label{eq:basis-param}
\end{equation}
with $\nu_0=(2.40\,\mathrm{fm})^{-2}$ and $b=1.25$,
irrespective of $(\ell,j)$.
Namely, 12 bases are employed for each $(\ell,j)$;
6 bases have real $\nu$ and the other 6 have complex $\nu$.
This set is quite similar to Set~C in Ref.~\cite{ref:Nak06}.

\subsection{Adaptability to wave functions with various size}

An appropriate set of the GEM bases
is capable of describing wave functions with various size.
This feature is desirable for self-consistent mean-field calculations
with deformation,
because in deformed nuclei the density distribution
depends on the direction,
and degree of the deformation is not known in advance.
We here show adaptability of the GEM
with respect to size of nuclei,
by presenting results of the spherical Hartree-Fock (HF) calculations.

\begin{figure}
\centerline{\includegraphics[scale=1.0]{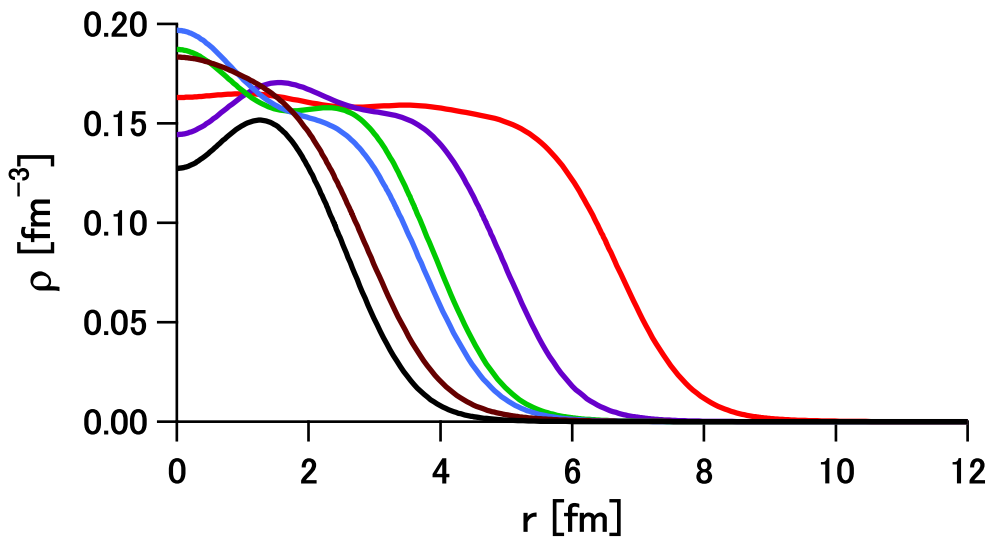}}
\vspace{2mm}
\caption{Density distribution of $^{16}$O (black line),
 $^{24}$O (brown line), $^{40}$Ca (blue line), $^{48}$Ca (green line),
 $^{90}$Zr (purple line) and $^{208}$Pb (red line),
 calculated by the GEM basis-set with the D1S interaction.
\label{fig:rho-sph}}
\end{figure}

Size of spherical nuclei,
which is typically represented by the rms radii,
depends on the mass number $A$,
apart from exotic structure such as neutron halos
near the drip line.
Many mean-field calculations have been implemented
by using the harmonic oscillator (HO) bases,
particularly when the effective interaction has finite ranges.
In the mean-field calculations with the HO bases,
the length parameter of the bases $b_\omega=1/\sqrt{M\omega}$
depends on $A$.
For stable nuclei, $b_\omega$ is almost proportional to $A^{1/6}$,
as $\omega\approx 41.2\,A^{-1/3}\,\mathrm{MeV}$,
although $b_\omega$ is often adjusted for individual nuclides
so as to minimize their energy $E$.
In contrast, since the GEM basis-set contains
Gaussians of various ranges,
even a single set can describe many nuclei
to good precision.
Binding energies and rms matter radii
calculated with the Gogny D1S interaction~\cite{ref:D1S}
are tabulated in Table~\ref{tab:sph-HF},
for the doubly-magic nuclei $^{16}$O, $^{24}$O, $^{40}$Ca, $^{48}$Ca,
$^{90}$Zr and $^{208}$Pb.
The values obtained from the GEM basis-set of Eq.~(\ref{eq:basis-param})
are compared with those from the $A$-dependent HO basis-set.
The Coulomb interaction between protons is handled
exactly~\cite{ref:NS02},
and the c.m. motion is fully removed
from the effective Hamiltonian before variation.
The influence of the c.m. motion on the rms matter radii
is treated in a similar manner~\cite{ref:Nak03}.
In the calculations using the HO bases,
the $b_\omega$ parameter of the bases is determined
from $\omega=41.2\,A^{-1/3}\,\mathrm{MeV}$,
and all the bases up to $N_\mathrm{osc}=15$ are included,
where $N_\mathrm{osc}$ is number of the oscillator quanta.
Because of the variational nature of the HF theory,
the lower energy indicates the more reliable result
for individual nuclei.
In this regard, the $A$-independent set of the GEM bases
gives no worse, even slightly better results
than the $A$-dependent HO basis-set except for $^{16}$O.
To show further the adaptability of the GEM basis-set
to nuclear size,
the calculated density distributions of the six nuclei are
illustrated in Fig.~\ref{fig:rho-sph}.

\begin{table}
\begin{center}
\caption{Binding energies $-E$ ($\mathrm{MeV}$)
 and rms matter radii $\sqrt{\langle r^2\rangle}$ ($\mathrm{fm}$)
 of $^{16}$O, $^{24}$O, $^{40}$Ca, $^{48}$Ca, $^{90}$Zr and $^{208}$Pb
 in the HF calculations with the D1S interaction.
 Results of the GEM bases and the HO bases are compared.
\label{tab:sph-HF}}
\begin{tabular}{ccrr}
\hline\hline
Nuclide && HO~~~& GEM~~\\ \hline
$^{16}$O & $-E$ & $129.638$ & $129.520$\\
 & $\sqrt{\langle r^2\rangle}$ & $2.603$ & $2.606$\\
$^{24}$O & $-E$ & $168.573$ & $168.598$\\
 & $\sqrt{\langle r^2\rangle}$ & $2.998$ & $3.014$\\
$^{40}$Ca & $-E$ & $344.470$ & $344.570$\\
 & $\sqrt{\langle r^2\rangle}$ & $3.370$ & $3.370$\\
$^{48}$Ca & $-E$ & $416.567$ & $416.764$\\
 & $\sqrt{\langle r^2\rangle}$ & $3.514$ & $3.513$\\
$^{90}$Zr & $-E$ & $785.126$ & $785.928$\\
 & $\sqrt{\langle r^2\rangle}$ & $4.238$ & $4.235$\\
$^{208}$Pb & $-E$ & $1638.094$ & $1639.047$\\
 & $\sqrt{\langle r^2\rangle}$ & $5.517$ & $5.513$\\
\hline\hline
\end{tabular}
\end{center}
\end{table}

In the results of Table~\ref{tab:sph-HF},
we use larger number of bases in the GEM calculation
than in the HO calculation.
Although one might think that this is not fair comparison,
it is not technically easy to increase the number of the HO bases,
because of round-off errors.
It is rather another advantage of the GEM
that we can avoid harmful round-off errors
even with a large number of bases.
Furthermore, it should be stressed that the GEM basis-set
used here is nucleus-independent,
demonstrating adaptability to nuclear size.
This yields an additional advantage in systematic calculations.
In the self-consistent mean-field calculations
such as HF and HFB with finite-range interactions,
computation of the two-body interaction matrix elements
is the most time-consuming part.
Since the basis-set of Subsec.~\ref{subsec:basis-set}
is applicable to many nuclei,
we do not have to repeat computation of the interaction matrix elements
once they are stored.

\subsection{Test on truncation with respect to $\ell$}

As viewed in Eq.~(\ref{eq:spwf}),
there is mixing of $\ell$ and $j$ in the s.p. wave functions
under the deformed mean field.
In particular, the $\ell$-mixing is relevant to the deformation.
The mixing of $j\,(=\ell\pm 1/2)$ is driven by the spin-orbit coupling.
In deformed mean-field calculations with the spherical bases,
we necessarily truncate the s.p. bases with respect to $\ell$
(and $j$),
by restricting the sum over $\ell$ in Eq.~(\ref{eq:spwf})
as $\ell\leq\ell_\mathrm{max}$.
It is an unavoidable question how large $\ell_\mathrm{max}$ is needed
for practical calculations.

To answer this question,
we consider an axially symmetric HO potential,
giving a s.p. Hamiltonian of
\begin{equation}
 \hat{h} = \frac{\mathbf{p}^2}{2M} + \frac{M}{2}\left[
 \omega_\perp^2(x^2+y^2)+\omega_z^2 z^2)\right]
 = \frac{\mathbf{p}^2}{2M} + \frac{M}{2} \omega_0^2\,r^2
 \left[1-\frac{4}{3}\sqrt{\frac{4\pi}{5}}\delta_\mathrm{def}\,
  Y^{(2)}_0(\hat{\mathbf{r}})\right]\,,
\label{eq:AHO}
\end{equation}
where $\omega_0^2=(2\omega_\perp^2+\omega_z^2)/3$
and $\delta_\mathrm{def}=(\omega_\perp^2-\omega_z^2)/2\omega_0^2$.
The exact eigenvalues of this Hamiltonian are obviously
$\varepsilon(n_\perp n_z)=\omega_\perp(n_\perp+1)+\omega_z(n_z+\frac{1}{2})$
($n_\perp, n_z= 0,1,\cdots$).
On the other hand, we obtain approximate eigenvalues
by diagonalizing this Hamiltonian with the spherical GEM bases,
by setting a certain value of $\ell_\mathrm{max}$.
Quality of the $\ell_\mathrm{max}$ truncation can be assessed
by comparing the approximate solutions to the exact ones.

\begin{figure}
\centerline{\includegraphics[width=13cm]{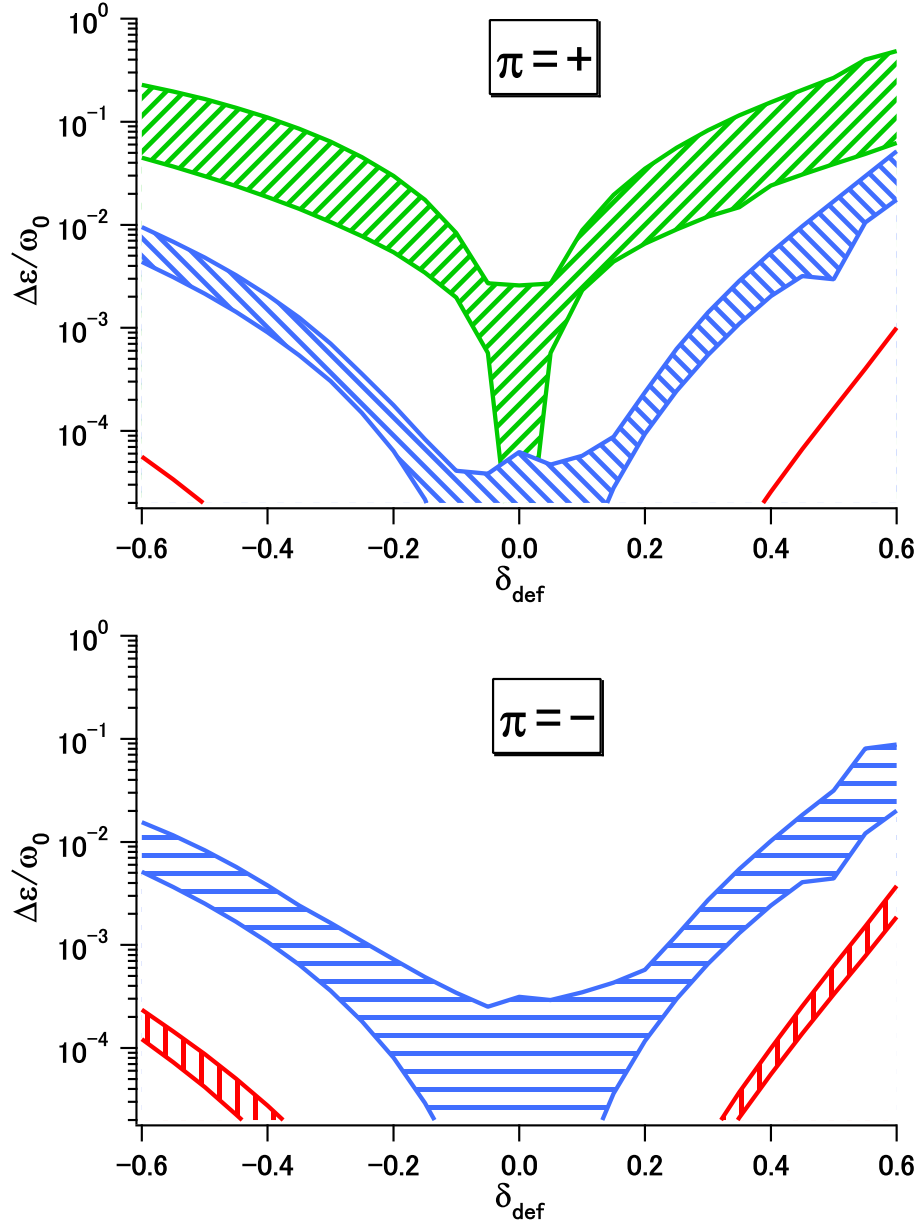}}
\vspace{2mm}
\caption{Errors of the GEM solutions in energy eigenvalues
 of the anisotropic harmonic oscillator
 of Eq.~(\protect\ref{eq:AHO}),
 whose ranges are represented by the hatched areas
 for each $N_\mathrm{osc}$ value.
 Upper panel: Errors for the $\pi=+$ levels.
 Red, blue and green areas show bounds of errors
 for the $N_\mathrm{osc}=0$, $2$ and $4$ levels, respectively.
 Lower panel: Errors for the $\pi=-$ levels.
 Red and blue areas represent bounds of errors
 for the $N_\mathrm{osc}=1$ and $3$ levels.
\label{fig:def-ho}}
\end{figure}

In Fig.~\ref{fig:def-ho}, we plot errors of the energy eigenvalues
obtained by the GEM,
$\mathit{\Delta}\varepsilon(n_\perp n_z)\equiv
\varepsilon_\mathrm{GEM}(n_\perp n_z)-
\varepsilon_\mathrm{exact}(n_\perp n_z)$,
as a function of $\delta_\mathrm{def}$,
taking $\ell_\mathrm{max}=7$ with the basis-set
of Eq.~(\ref{eq:basis-param}).
Although we set $(\omega_\perp^2 \omega_z)^{1/3}
=41.2\,A^{-1/3}\,\mathrm{MeV}$
with $A=24$ at each $\delta_\mathrm{def}$,
the errors are insensitive to $\omega_0$
if we measure $\mathit{\Delta}\varepsilon$ in unit of $\omega_0$.
We have confirmed that the results presented in Fig.~\ref{fig:def-ho}
are almost unchanged even for $A=208$.
The eigenvalues belonging to equal $N_\mathrm{osc}(=n_\perp+n_z)$
have errors of the same orders of magnitude
(except at $\delta_\mathrm{def}=0$).
This is because the levels belonging to equal $N_\mathrm{osc}$
significantly mix one another owing to their close energies,
while admixture of the $N_\mathrm{osc}\pm 2$ levels
occurs only perturbatively.
We therefore classify the s.p. levels by $N_\mathrm{osc}$
and show their errors
as a function of $\delta_\mathrm{def}$ by the hatched areas,
which are edged by the minimum and maximum errors
for each $N_\mathrm{osc}$.

At $\delta_\mathrm{def}=0$ (\textit{i.e.} $\omega_\perp=\omega_z$),
the energy eigenvalues become $\varepsilon(n_\perp n_z)
= \omega_0 (N_\mathrm{osc}+\frac{3}{2})$.
The errors at $\delta_\mathrm{def}=0$ exclusively come from
the choice of the radial parameters in Eq.~(\ref{eq:basis-param}),
irrelevant to the $\ell$-mixing.
We have much smaller errors at $\delta_\mathrm{def}=0$
than those at finite $\delta_\mathrm{def}$.
Because the radial parts are well described,
the errors at $\delta_\mathrm{def}\ne 0$ predominantly depend on
how well the $\ell$-mixing is taken into account.
Irregularities at $\delta_\mathrm{def}\geq 0.5$ as well as
at $\delta_\mathrm{def}=0$ occur due to crossing of the levels
that have equal $\pi$ and $m$ quantum numbers.

Whereas it depends on the situation
what precision is required in the calculation,
we here assume the criterion to be
$\mathit{\Delta}\varepsilon\lesssim 0.01\,\omega_0$,
which corresponds to
$\mathit{\Delta}\varepsilon\lesssim 0.1\,\mathrm{MeV}$
for nuclei with $\omega_0\sim 10\,\mathrm{MeV}$.
With the $\ell_\mathrm{max}=7$ truncation,
all the $N_\mathrm{osc}\leq 3$ levels satisfy this criterion
at $-0.5\lesssim\delta_\mathrm{def}\lesssim +0.35$.
This region of $\delta_\mathrm{def}$ covers
deformation of most nuclei,
as long as their low-energy states are concerned.
Since $\mathit{\Delta}\varepsilon$ exceeds $0.01\,\omega_0$
even at $\delta_\mathrm{def}\approx 0.1$,
we cannot expect that the $\ell_\mathrm{max}=7$ truncation
works well for the $N_\mathrm{osc}\geq 4$ levels.
It should be noticed that the crossing between
the highest level of $N_\mathrm{osc}=3$
and the lowest level of $N_\mathrm{osc}=5$
takes place at $\delta_\mathrm{def}=+0.41$ and $-0.56$.
These crossing points are close to the edges of the region
where the above criterion is fulfilled.
Beyond these $\delta_\mathrm{def}$ values
some of the $N_\mathrm{osc}=5$ levels,
for which we need higher $\ell_\mathrm{max}$, are occupied
before all the $N_\mathrm{osc}=3$ levels are,
although the crossing levels do not necessarily mix
since they may have different quantum numbers $m$.
Similar crossing between the $N_\mathrm{osc}=4$ and $6$ levels
occurs at $\delta_\mathrm{def}\approx +0.34$ and $-0.44$.
Thus the crossing points between $N_\mathrm{osc}$ and $N_\mathrm{osc}+2$
levels are not very sensitive to $N_\mathrm{osc}$.
Although the spin-orbit splitting influences the level crossing points,
the current argument will hold in more realistic cases to a good degree,
because the spin-orbit splitting is much smaller
than $2\omega_0$.
Thus it will be reasonable to state that
we should take $\ell_\mathrm{max}\geq N_\mathrm{osc}^F+4$
in deformed mean-field calculations,
where $N_\mathrm{osc}^F$ is defined to be
the highest oscillator quantum number for the s.p. levels under interest.
If we take $\ell_\mathrm{max}\geq N_\mathrm{osc}^F+6$,
we can expect better precision by about two orders of magnitude;
the error on the level corresponding to $N_\mathrm{osc}^F$
will be $\mathit{\Delta}\varepsilon^F\lesssim 10^{-4}\,\omega_0$
for low energy states.

While we have discussed errors on the s.p. energies,
mean-field calculations are carried out
by minimizing the total energy of each nucleus.
Because the errors grow for the larger $|\delta_\mathrm{def}|$,
calculations with the spherical bases tend to yield
higher energy for deformed states than for spherical states.
We put $N_\mathrm{osc}^F$
to be the value for the Fermi level,
which can be estimated as $N_\mathrm{osc}^F\approx (3A/2)^{1/3}$~\cite{BM1}.
The number of nucleons occupying the $N_\mathrm{osc}^F$ levels
should be $2(N_\mathrm{osc}^F)^2\approx (3\sqrt{2}A)^{2/3}$.
Dominated by the errors on the $N_\mathrm{osc}^F$ levels,
error on the total energy due to the $\ell_\mathrm{max}$ truncation
is estimated to be $\mathit{\Delta}E\approx
(3\sqrt{2}A)^{2/3}\,\mathit{\Delta}\varepsilon^F$.
Inserting $\mathit{\Delta}\varepsilon^F\approx 0.01\,\omega_0$
and $\omega_0\approx 40\,\mathrm{MeV}$,
we obtain $\mathit{\Delta}E\approx A^{1/3}\,\mathrm{MeV}$
for well-deformed configurations,
while this error is absent in spherical configurations.
We should keep this point in mind
when comparing energies of a spherical minimum and a deformed one.

\section{Implementation of self-consistent mean-field calculations
  with deformation}
\label{sec:impl}

We have discussed that
the spherical GEM bases shown in Subsec.~\ref{subsec:basis-set}
is promising in describing deformed nuclei.
We next present how we implement self-consistent HF or HFB calculations
using the GEM bases.

One of the advantages of the GEM algorithm
in Refs.~\cite{ref:NS02,ref:Nak06}
was that finite-range effective interactions
are tractable~\cite{ref:Nak03}
by computing two-body interaction matrix elements of the interactions,
even for the LS and the tensor channels.
This advantage is maintained in the deformed cases.
While in the spherical cases
the number of necessary matrix elements is reduced to great extent
due to the symmetry of the one-body fields,
we need all the non-vanishing two-body matrix elements
in the deformed mean-field calculations.
Since we adopt the spherical GEM bases,
the matrix elements can straightforwardly be computed
according to the formulae shown in Refs.~\cite{ref:NS02,ref:Nak06}.
It costs 4.3\,GB of memory or disk
to store the matrix elements with $\ell\leq 7$ in double precision.
We handle the Coulomb interaction in a similar manner,
which needs additional 2.2\,GB.
Although it is time-consuming task to compute them,
those matrix elements are useful for various calculations
once we store them;
not only for HF and HFB calculations of many nuclei,
but also for calculations in the random-phase approximation,
as will be discussed elsewhere.

The density-dependent interaction should be renewed at each iteration.
This is not a difficult task as long as this part of the interaction
has a contact form as in Ref.~\cite{ref:Nak03}.

Folding the stored matrix elements with given s.p. wave functions,
we construct the s.p. Hamiltonian
that preserves the $(\pi,m)$ quantum numbers.
The s.p. wave functions are then obtained
by solving the HF or HFB equation
as a generalized eigenvalue problem, for each $(\pi,m)$.
Starting from appropriate initial values,
we repeat this procedure iteratively until convergence.
Alternatively, we can apply the gradient method
to obtain the energy minimum by using the s.p. Hamiltonian.

The present algorithm keeps the advantages of the GEM
in the spherical mean-field calculations.
We here list them again:
(i) it is efficient in describing
the energy-dependent asymptotics of s.p. wave functions
at large $r$,
(ii) we can handle various effective interactions,
including those having non-locality,
and (iii) the basis parameters are insensitive to nuclide,
thereby a single-set of bases is applicable
to wide mass range of nuclei.

\section{Numerical examples for magnesium isotopes}
\label{sec:test}

We now apply the present method of the deformed HF and HFB calculations
to actual nuclei.
We take even-$N$ magnesium isotopes as examples,
which contain a well-deformed stable nucleus $^{24}$Mg,
several nuclei on the island of inversion
such as $^{32}$Mg and $^{34}$Mg,
and $^{40}$Mg for which the $N=28$ magicity has been predicted
to disappear in several calculations.
We exactly treat the Coulomb energy~\cite{ref:NS02},
and both the one- and two-body terms of the c.m. part are subtracted
from the Hamiltonian before variation,
unless mentioned explicitly.
The GEM bases of Subsec.~\ref{subsec:basis-set} are employed
with the $\ell_\mathrm{max}=7$ truncation.
This fulfills $\ell_\mathrm{max}\geq N_\mathrm{osc}^F+4$
for the ground states of all the Mg nuclei to be presented,
since the last neutron occupies an $N_\mathrm{osc}=3$ orbit
in the neutron-rich Mg nuclei.

\subsection{Comparison with previous calculations}

We examine how well the present method can describe
the deformed Mg nuclei.
We first show results of the HF calculation
with the D1 parameter-set~\cite{ref:Gogny} of the Gogny interaction.

In Ref.~\cite{ref:HF-D1},
an axial HF calculation using the HO bases
has been implemented for many Mg isotopes.
Although the basis-set consists only of $N_\mathrm{osc}\leq 4$,
the oscillator length $b_\omega$ is optimized
except for $^{36}$Mg and $^{38}$Mg.
In Table~\ref{tab:HF_Mg}
we compare the present GEM results for the Mg isotopes
with those in Ref.~\cite{ref:HF-D1}.
For $^{22,24,28}$Mg, binding energies
computed in the antisymmetrized molecular dynamics (AMD)~\cite{ref:AMD-Mg}
are reported.
The AMD is a powerful tool to study structure
of light to medium-mass nuclei.
In the AMD approach the total wave function is represented
by a Slater determinant of Gaussian wave packets
of constituent nucleons.
While ranges of the Gaussians are taken to be equal
for all nucleons in the nucleus,
the nucleons have different central positions from one another
and the positions are optimized
without assuming the axial symmetry.
Whereas the AMD has been extended
by incorporating the projections
and superposing many Slater determinants,
we here use the results of its simplest version for comparison,
because it is analogous to the HF approximation.

With respect to the spurious c.m. motion,
it is popular in the mean-field calculations so far
to remove only the one-body term before variation.
One of the present GEM results (denoted by GEM$_1$)
are obtained by this prescription.
While influence of the spurious c.m. motion
can be fully removed in the AMD,
only the one-body term is subtracted before variation
in the AMD results shown in Table~\ref{tab:HF_Mg}.
Therefore the GEM$_1$ values should be compared
to the HO and the AMD ones.
We additionally present the values
in which the two-body term of the c.m. Hamiltonian is also subtracted
before variation in the GEM calculation,
and denote them by GEM$_2$.

\begin{table}
\begin{center}
\caption{Comparison of the axial HF results
 for binding energies $-E$ ($\mathrm{MeV}$)
 and intrinsic mass quadrupole moments $Q_0$ ($\mathrm{fm}^2$)
 of the Mg isotopes.
 The present GEM results are compared with
 those of the HO bases~\protect\cite{ref:HF-D1} and
 the AMD calculation~\protect\cite{ref:AMD-Mg}.
 The one-body c.m. energy is subtracted in GEM$_1$
 as well as in the HO and the AMD results,
 while the two-body c.m. energy is also removed in GEM$_2$.
 The D1 interaction is adopted for all the calculations.
\label{tab:HF_Mg}}
\begin{tabular}{ccrrrr}
\hline\hline
Nuclide && HO~~~& AMD~~& GEM$_1$~& GEM$_2$~\\ \hline
$^{22}$Mg & $-E$ & --- & $161.26$ & $169.03$ & $162.44$ \\
 & $Q_0$ & --- & --- & $100.84$ & $104.53$ \\
$^{24}$Mg & $-E$ & $193.90$ & $188.33$ & $195.73$ & $188.74$ \\
 & $Q_0$ & $113.95$ & --- & $114.39$ & $118.11$ \\
$^{26}$Mg & $-E$ & $211.11$ & --- & $212.99$ & $205.29$ \\
 & $Q_0$ & $76.51$ & --- & $78.20$ & $81.09$ \\
$^{28}$Mg & $-E$ & $227.82$ & $223.07$ & $230.14$ & $222.45$ \\
 & $Q_0$ & $101.37$ & --- & $101.41$ & $105.06$ \\
$^{30}$Mg & $-E$ & $237.69$ & --- & $240.34$ & $232.36$ \\
 & $Q_0$ & $76.10$ & --- & $75.07$ & $77.29$ \\
$^{32}$Mg & $-E$ & $245.98$ & --- & $249.05$ & $240.97$ \\
 & $Q_0$ & $34.27$ & --- & $37.13$ & $38.33$ \\
$^{34}$Mg & $-E$ & $249.94$ & --- & $254.77$ & $246.43$ \\
 & $Q_0$ & $111.70$ & --- & $128.36$ & $131.92$ \\
$^{36}$Mg & $-E$ & $253.79$ & --- & $260.49$ & $251.84$ \\
 & $Q_0$ & $146.17$ & --- & $176.50$ & $180.81$ \\
$^{38}$Mg & $-E$ & $251.69$ & --- & $261.69$ & $252.62$ \\
 & $Q_0$ & $168.65$ & --- & $175.06$ & $179.47$ \\
\hline\hline
\end{tabular}
\end{center}
\end{table}

Since the HF approximation holds variational nature,
the lower ground-state energy indicates the more reliable result
(\textit{i.e.} the closer to the true minimum).
In this respect the present GEM bases give more favorable results
than the HO and the AMD bases.
The present method gives $7-8\,\mathrm{MeV}$ lower energies
than the AMD in $^{22,24,28}$Mg.
Compared to the HO calculation in Ref.~\cite{ref:HF-D1},
the energy gain in the present calculation grows
as going to the neutron-rich region,
$\sim 2\,\mathrm{MeV}$ in $^{24}$Mg to $\sim 10\,\mathrm{MeV}$
in $^{38}$Mg.
It may be due to spatially broad distribution of density
in the neutron-rich region,
which is well described by the GEM
but is hard to be reproduced with the HO bases.
It should be noticed that $^{38}$Mg is unbound
in the result of Ref.~\cite{ref:HF-D1}
because it has higher energy than $^{36}$Mg,
while it is bound in the present calculation.
This exemplifies importance of numerical algorithm
that appropriately handles spatial extension of wave functions
when investigating the neutron drip line.

The two-body term of the c.m. Hamiltonian
affects the energies to sizable amount,
as clearly viewed in Table~\ref{tab:HF_Mg}.
Difference between the GEM$_1$ and GEM$_2$ energies slightly grows
for increasing $A$.

We next turn to the HFB calculations.
We compare the GEM results for $^{30-34}$Mg
with those obtained from the HO bases
truncated by $N_\mathrm{osc}\leq 10$~\cite{ref:HO-Mg},
in Table~\ref{tab:Mg30-34}.
The D1S parameter-set~\cite{ref:D1S} of the Gogny interaction is adopted,
and the c.m. Hamiltonian is fully subtracted before iteration
in both calculations.
In Ref.~\cite{ref:HO-Mg},
the spherical HO bases are used,
so that the angular-momentum projection could be carried out afterward.
The exchange term of the Coulomb energy is handled
in the Slater approximation in Ref.~\cite{ref:HO-Mg}.

\begin{table}
\begin{center}
\caption{Binding energies $-E$ (MeV) of $^{30,32,34}$Mg
 in the HFB calculations with the D1S interaction.
 The GEM results are compared with those of the HO bases
 in Ref.~\protect\cite{ref:HO-Mg}.
\label{tab:Mg30-34}}
\begin{tabular}{crrr}
\hline\hline
Nuclide & HO~~~~& GEM~~~& Exp.~~~\\
\hline
$^{30}$Mg & $239.30~~$ & $239.47~~$ & $241.63~~$ \\
$^{32}$Mg & $248.22~~$ & $248.30~~$ & $249.69~~$\\
$^{34}$Mg & $252.82~~$ & $254.02~~$ & $256.59~~$ \\
\hline\hline
\end{tabular}
\end{center}
\end{table}

We find that the present method gives
lower energies in all of $^{30-34}$Mg
than the HO calculation in Ref.~\cite{ref:HO-Mg}.
If we take differences in computational procedure
irrelevant to the basis-sets
(\textit{e.g.} the treatment of the Coulomb exchange term)
into consideration,
it is fair to say that the present results are no worse
than the HO ones in Ref.~\cite{ref:HO-Mg}
from the variational viewpoint.
We obtain lower energy in the present calculation
than in Ref.~\cite{ref:HO-Mg} by greater than $1\,\mathrm{MeV}$
for the unstable nucleus $^{34}$Mg,
while by less than $0.2\,\mathrm{MeV}$ for $^{30,32}$Mg.
This suggests that the spatial extention
is more important than the mixing of the $\ell>7$ components
particularly in $^{34}$Mg.
Notice that the HO basis-set in Ref.~\cite{ref:HO-Mg}
contains up to $\ell=10$.

\subsection{Neck structure of neutron halo in $^{40}$Mg}

As we have pointed out, the GEM can describe
the wave-function asymptotics at large $r$
to reasonable precision.
Taking this advantage, we investigate density distribution
of axially deformed drip-line nuclei,
within the mean-field approximation.

We here discuss in the HF framework.
The s.p. wave function $\psi_{n\pi m}(\mathbf{r})$ in Eq.~(\ref{eq:spwf})
can be decomposed by a sum of spherical wave functions,
\begin{equation}
 \psi_{n\pi m}(\mathbf{r}) = \sum_{\ell\sigma}
  \bar{c}^{(n)}_{\ell\sigma m}\,
  \bar{\varphi}^{(n)}_{\ell\sigma m}(\mathbf{r})\,.
\end{equation}
Here $\bar{\varphi}^{(n)}_{\ell\sigma m}(\mathbf{r})
=\bar{R}^{(n)}_{\ell\sigma m}(r)\,Y^{(\ell)}_{m_\ell}(\hat{\mathbf{r}})
\,\chi_\sigma$ with $m_\ell+\sigma/2=m$,
which is defined by the $(\ell\sigma)$-projection
on $\psi_{n\pi m}(\mathbf{r})$ besides a constant factor.
The nuclear force becomes negligible at sufficiently large $r$.
There the coordinate-represented HF Hamiltonian for neutrons
becomes approximately spherically symmetric
and the components having different $\ell$ decouple
to one another~\cite{ref:MNA97}.
The spin-orbit coupling, which is an effect of the nuclear force,
also becomes negligible,
and thereby the $\sigma$ (and $m_\ell$) value is frozen.
After each $(\ell\sigma)$ component propagates
over a certain region of $r$,
$\psi_{n\pi m}(\mathbf{r})$ will be dominated
by the $\bar{\varphi}^{(n)}_{\ell\sigma m}(\mathbf{r})$ component
of the lowest possible $\ell$
(\textit{i.e.} $\ell=|m|\pm 1/2$, with the sign fixed by the parity)
due to the centrifugal barrier~\cite{ref:MNA97,ref:Ham04},
although the degree of the dominance
depends on characters of the deformed orbit.
The asymptotic form of $\bar{R}^{(n)}_{\ell\sigma m}(r)$ at large $r$
is $e^{-\eta_{n\pi m} r}/r$
with $\eta_{n\pi m}=\sqrt{2M|\varepsilon_{n\pi m}|}$
($\varepsilon_{n\pi m}$ is the s.p. energy).
Hence the neutron density distribution
asymptotically behaves as
\begin{equation}
 \rho_n(\mathbf{r}) \approx g_{\!F}\,|\psi_{\!F}(\mathbf{r})|^2
  \approx g_{\!F}\,\frac{e^{-2\eta_{\!F} r}}{r^2}
  \sum_\sigma |\bar{c}^{(n_{\!F})}_{\ell_{\!F}\,\sigma\,m_{\!F}}|^2\,
  \big|Y^{(\ell_{\!F})}_{m_{\!F}-\sigma/2}(\hat{\mathbf{r}})\big|^2\,,
 \label{eq:asymp}
\end{equation}
where the subscript $F$ represents the highest occupied level,
$\ell_{\!F}$ is determined by $\ell_{\!F}=|m_{\!F}|\pm 1/2$
and $(-)^{\ell_{\!F}}=\pi_{\!F}$,
and $g_{\!F}$ stands for the occupation number on the level $F$.
We hereafter assume $m_{\!F}>0$ without loss of generality,
owing to the reflection symmetry.

In connection to halos,
the $\ell_{\!F}=0$ and $1$ cases are particularly interesting.
For $\ell_{\!F}=0$, which implicates $\pi_{\!F}=+$ and $m_{\!F}=1/2$,
Eq.~(\ref{eq:asymp}) is reduced to
\begin{equation}
 \rho_n(\mathbf{r})
  \approx g_{\!F}\,\frac{e^{-2\eta_{\!F} r}}{r^2}\,
  |\bar{c}^{(n_{\!F})}_{0\,+\,1/2}|^2\,
  \big|Y^{(0)}_0(\hat{\mathbf{r}})\big|^2
  = \frac{g_{\!F}}{4\pi}\,\frac{e^{-2\eta_{\!F} r}}{r^2}\,
  |\bar{c}^{(n_{\!F})}_{0\,+\,1/2}|^2\,,
 \label{eq:asymp0}
\end{equation}
giving isotropic density in the asymptotic region.
Hence halos formed by $\pi=+$ last nucleon are spherically symmetric,
as long as the $\ell\geq 2$ components can be neglected.
For $\ell_{\!F}=1$ which presumes $\pi_{\!F}=-$,
there are two possibilities $m_{\!F}=1/2$ and $3/2$.
In the $m_{\!F}=3/2$ case, Eq.~(\ref{eq:asymp}) becomes
\begin{equation}
 \rho_n(\mathbf{r})
  \approx g_{\!F}\,\frac{e^{-2\eta_{\!F} r}}{r^2}\,
  |\bar{c}^{(n_{\!F})}_{1\,+\,3/2}|^2\,
  \big|Y^{(1)}_1(\hat{\mathbf{r}})\big|^2
  = \frac{3}{4\pi}g_{\!F}\,\frac{e^{-2\eta_{\!F} r}}{r^2}\,
  |\bar{c}^{(n_{\!F})}_{1\,+\,3/2}|^2\,
  \frac{x^2+y^2}{2r^2}\,.
 \label{eq:asymp1a}
\end{equation}
This asymptotic component is obviously deformed,
having vanishing contribution in the $z$ direction.
In the $m_{\!F}=1/2$ case, we obtain
\begin{eqnarray}
 \rho_n(\mathbf{r})
  &\approx& g_{\!F}\,\frac{e^{-2\eta_{\!F} r}}{r^2}\,
  \left(|\bar{c}^{(n_{\!F})}_{1\,+\,1/2}|^2\,
  \big|Y^{(1)}_0(\hat{\mathbf{r}})\big|^2
  +|\bar{c}^{(n_{\!F})}_{1\,-\,1/2}|^2\,
  \big|Y^{(1)}_1(\hat{\mathbf{r}})\big|^2\right)
  \nonumber\\
  &=& \frac{3}{4\pi}g_{\!F}\,\frac{e^{-2\eta_{\!F} r}}{r^2}\,
  \left(|\bar{c}^{(n_{\!F})}_{1\,+\,1/2}|^2\,\frac{z^2}{r^2}
  +|\bar{c}^{(n_{\!F})}_{1\,-\,1/2}|^2\,
  \frac{x^2+y^2}{2r^2}\right) \nonumber\\
  &=& \frac{g_{\!F}}{4\pi}\,\frac{e^{-2\eta_{\!F} r}}{r^2}\,
  \left[\big(|\bar{c}^{(n_{\!F})}_{1\,+\,1/2}|^2
   +|\bar{c}^{(n_{\!F})}_{1\,-\,1/2}|^2\big)
  +\sqrt{\frac{4\pi}{5}}\big(2|\bar{c}^{(n_{\!F})}_{1\,+\,1/2}|^2
  -|\bar{c}^{(n_{\!F})}_{1\,-\,1/2}|^2\big)\,
  Y^{(2)}_0(\hat{\mathbf{r}})\right] \,. \nonumber\\
 \label{eq:asymp1b}
\end{eqnarray}
The asymptotic component of Eq.~(\ref{eq:asymp1b}) can also be deformed,
and its degree depends on $|\bar{c}^{(n_{\!F})}_{1\,+\,1/2}|$
and $|\bar{c}^{(n_{\!F})}_{1\,-\,1/2}|$.
Thus, within the HF approximation,
a deformed halo is expected in axially deformed drip-line nuclei
with the last nucleon occupying a $(\pi,m)=(-,1/2)$ or $(-,3/2)$ orbit.

Connected to the Nilsson model,
the asymptotic quantum numbers have widely been used
in description of well-deformed nuclei,
comprising $m_\ell$ (denoted by $\Lambda$ in the Nilsson model).
Therefore $\sigma$ becomes an approximate quantum number.
If this is the case,
one of $|\bar{c}^{(n_{\!F})}_{1\,\pm\,1/2}|$ is much greater
than the other in Eq.~(\ref{eq:asymp1b}).
Then the asymptotic behavior of $\rho_n(\mathbf{r})$
depends largely on direction
in the $(\pi_{\!F},m_{\!F})=(-,1/2)$ case.
In the extreme case where one of $\bar{c}^{(n_{\!F})}_{1\,\pm\,1/2}$
vanishes, the damping factor $e^{-2\eta_{\!F} r}/r^2$
is missing in a certain direction;
if $\bar{c}^{(n_{\!F})}_{1\,-\,1/2}=0$,
$\rho_n$ damps according to the $e^{-2\eta_{\!F} r}/r^2$ factor
in the $z$ direction,
but damps faster in the $x$ and $y$ directions.
The absence of the $e^{-2\eta_{\!F} r}/r^2$ asymptotic component
is significant only in the vicinity of the $xy$ plane.
Hence a neck structure of a halo
is expected for drip-line nuclei with $(\pi_{\!F},m_{\!F})=(-,1/2)$
that is dominated by the $m_\ell=0$ component.
In contrast, the density in nuclei with $(\pi_{\!F},m_{\!F})=(-,3/2)$
or $(-,1/2)$ dominated by $m_\ell=1$
may have dips on the $z$ axis,
forming a halo with an apple-like shape.

In the Mg isotopes,
the neutron-drip line has been predicted to lie at $N=28$
in several mean-field calculations so far~\cite{ref:SHFB-Mg,ref:HO-Mg2}.
In a recent experiment $^{40}$Mg has been confirmed
to be bound~\cite{ref:Mg40-exp}.
We obtain the drip line at $N=28$,
as reported in Ref.~\cite{ref:HO-Mg2};
$^{40}$Mg is bound while $^{42}$Mg has higher energy than $^{40}$Mg.
In the present HF result $^{40}$Mg has a prolate shape
at its energy minimum, with $E=-260.7\,\mathrm{MeV}$,
$\sqrt{\langle r^2\rangle}=3.63\,\mathrm{fm}$
and $Q_0=227\,\mathrm{fm}^2$.
It is noted that this energy is appreciably lower
than in the $N_\mathrm{osc}\leq 17$ HO calculation~\cite{ref:HO-Mg2}.
This is probably connected to the spatial extension
of the wave function shown below.
The last two neutrons,
which have $\varepsilon_{\!F}=-1.14\,\mathrm{MeV}$,
occupy a $(\pi,m)=(-,1/2)$ orbit
that has large portion of the $\ell=1$ component by $68\%$.
Moreover, we have $|\bar{c}^{(n_{\!F})}_{1\,+\,1/2}|^2\big/
\big(|\bar{c}^{(n_{\!F})}_{1\,+\,1/2}|^2+|\bar{c}^{(n_{\!F})}_{1\,-\,1/2}|^2\big)
\approx 0.95$ for this orbit.
Applying the above argument,
it is likely that $\rho_n(\mathbf{r})$ of $^{40}$Mg has a halo
oriented to the $z$ direction,
but this halo component is almost missing on the $xy$ plane,
leading to neck structure of the neutron halo.

\begin{figure}
\centerline{\includegraphics[scale=1.0]{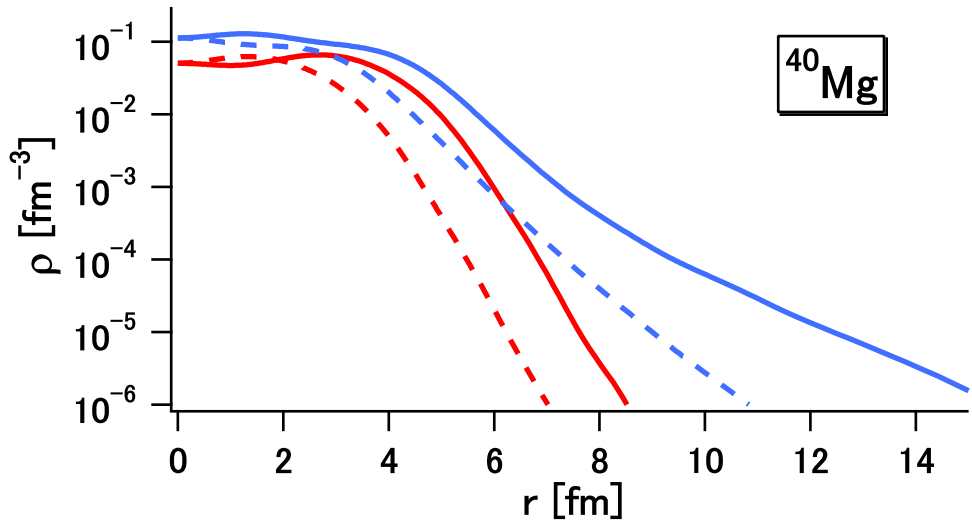}}
\vspace{2mm}
\caption{Proton (in red color) and neutron (in blue color) densities
 of $^{40}$Mg in the $x$ (dashed lines) and the $z$ (solid lines) directions,
 calculated with the D1S interaction.
\label{fig:rho-Mg40}}
\end{figure}

In Fig.~\ref{fig:rho-Mg40},
$\rho_\tau(x=y=0,z)$ and $\rho_\tau(x,y=z=0)$ $(\tau=p,n)$
of $^{40}$Mg are depicted.
Although $\rho_p(z)$ is elongated in comparison to $\rho_p(x)$
to certain extent because of the prolate deformation,
they damp with nearly equal slope at $r\gtrsim 5\,\mathrm{fm}$.
We find halo structure in $\rho_n(z)$ at $z\gtrsim 8\,\mathrm{fm}$
with the asymptotic behavior that is consistent with $\varepsilon_{\!F}$
to good approximation.
It is remarked that the component with the same asymptotics
is highly suppressed in the $x$ direction.

\begin{figure}
\centerline{\includegraphics[scale=0.45]{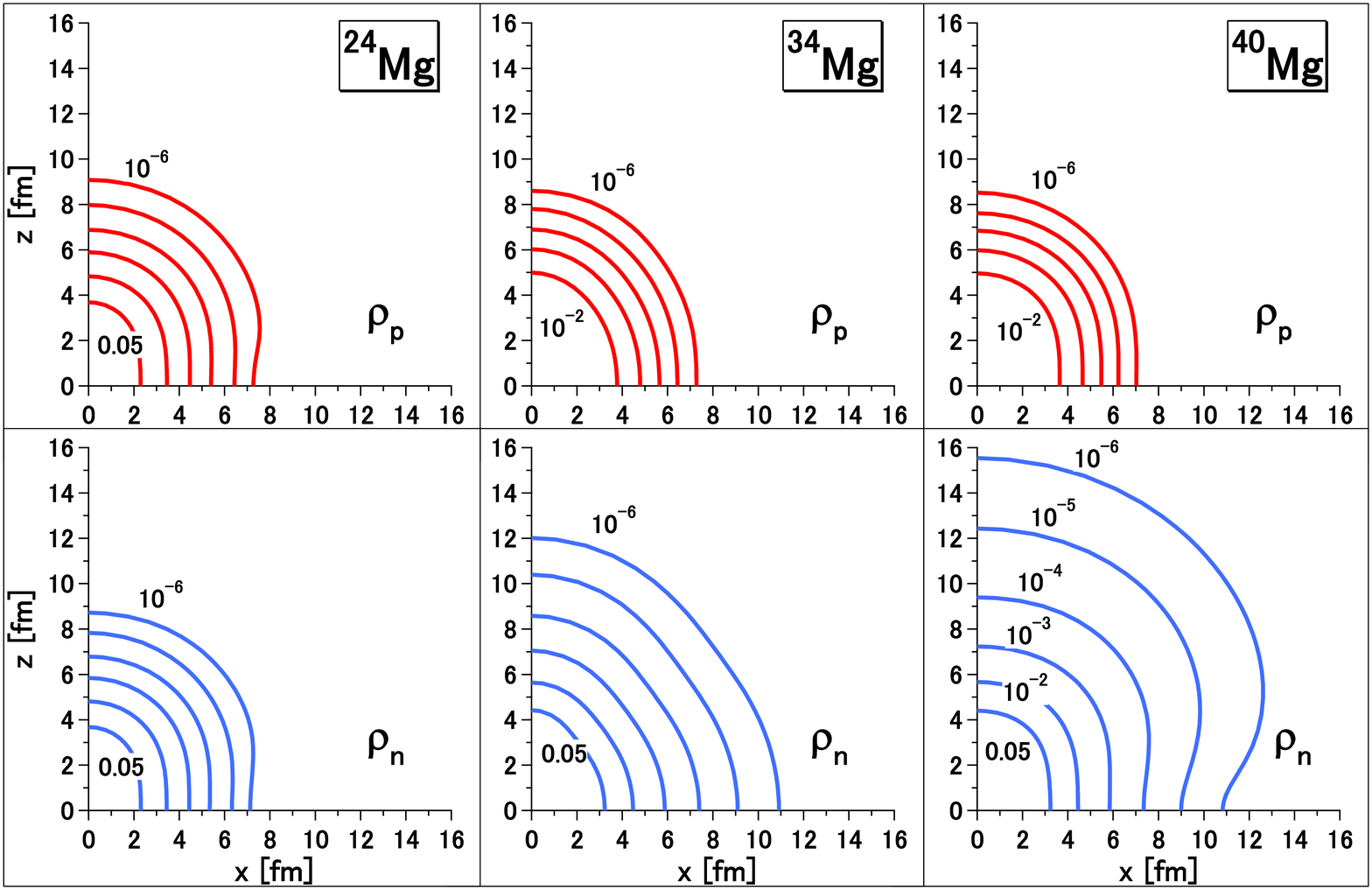}}
\vspace{2mm}
\caption{Contour plot of $\rho_\tau$ ($\tau=p,n$) on the $xz$-plane
 for $^{24}$Mg, $^{34}$Mg and $^{40}$Mg,
 calculated with the D1S interaction.
 Positions of $\rho_\tau=0.05$, $10^{-2}$, $10^{-3}$, $10^{-4}$,
 $10^{-5}$ and $10^{-6}\,\mathrm{fm}^{-3}$ are presented.
\label{fig:rhcnt-Mg}}
\end{figure}

To observe the shape of $^{40}$Mg at low density,
equi-density lines on the $xz$-plane
are drawn in Fig.~\ref{fig:rhcnt-Mg}.
For comparison, similar plots for $^{24}$Mg and $^{34}$Mg
are given as well.
All the three Mg nuclei have prolate deformation
in the present HF calculation.
The equi-density lines distribute with almost equal intervals
from $10^{-2}$ to $10^{-6}\,\mathrm{fm}^{-3}$,
indicating exponential decrease of $\rho_\tau$.
With the exception of $\rho_n$ in $^{40}$Mg,
the size of the intervals does not depend on direction.
Although the $\rho_\tau=10^{-6}\,\mathrm{fm}^{-3}$ line in $^{24}$Mg
is slightly constricted in the $x$ direction,
it could be influenced by deviation from the exponential asymptotics
due to numerical errors.
The contour plot of $\rho_n$ in $^{40}$Mg is intriguing.
The large intervals of the equi-density lines correspond to
the slow damping of $\rho_n$ due to the small $\varepsilon_{\!F}$.
However, the intervals are comparable
to those in $^{34}$Mg in the $x$ direction.
We view the neck structure in the low $\rho_n$ region.

Because it mixes the $m_\ell=0$ and $1$ components,
the pair correlation tends to restore the spherical symmetry
in the density distribution in the asymptotic region.
Depending on degree of the mixing,
the neck structure may survive or disappear.
However, in the present HFB calculation with the D1S interaction,
we do not find a superfluid solution
having lower energy than the HF minimum in $^{40}$Mg.
Thus the density distribution of $^{40}$Mg
shown in Figs.~\ref{fig:rho-Mg40} and \ref{fig:rhcnt-Mg} is not altered.

If the $\ell\geq 2$ components remain sizable
in the asymptotic region,
deformed halos (including neck structure) with other $(\pi,m)$
may be present.
In Ref.~\cite{ref:def-halo},
possibility of deformed halos depending on the direction
was pointed out for $^{11}$Be and $^{13}$C,
based on the Skyrme HFB calculations.
Those halos are formed by a $(\pi,m)=(+,1/2)$ orbit.

There remain problems with respect to deformed halos:
(a) how correlations beyond HFB
(including restoration of the rotational symmetry) affect them,
and (b) whether they are detectable in experiments.
Although both questions are very important,
it is not easy to give satisfactory answers at this moment
and we leave these problems to future studies.
We here emphasize that such exotic structure of nuclear halos
can be investigated only via numerical methods
that are capable of describing the wave-function asymptotics appropriately.

\section{Summary}
\label{sec:summary}

We extensively develop a new method of implementing
the Hartree-Fock (HF) and the Hartree-Fock-Bogolyubov (HFB) calculations
of nuclei with deformation,
applying the Gaussian expansion method (GEM).
Owing to the adaptability in describing radial degrees of freedom,
which is confirmed by the HF calculations from $^{16}$O to $^{208}$Pb,
we adopt the spherical GEM bases.
We argue how large $\ell$ should be taken into account,
by comparing the numerical solutions
obtained from the spherical GEM bases with the analytic ones
in the axially deformed harmonic oscillator.
The present method maintains three notable advantages
of the GEM algorithm for the mean-field calculations:
(i) we can efficiently describe
the energy-dependent asymptotics of single-particle (s.p.) wave functions
at large $r$,
(ii) we can handle various effective interactions,
including those having non-locality,
and (iii) a single-set of bases is applicable
to wide mass range of nuclei
and therefore is suitable to systematic calculations.

The present method is applied to magnesium nuclei
with the Gogny force.
We show that the present results are no worse
than those in the literatures, from the variational viewpoint.
Compared with the conventional method using the harmonic oscillator bases,
the present method is suitable particularly
to nuclei far from the $\beta$ stability.
For $^{40}$Mg, we suggest neck structure of a neutron halo,
which arises due to the asymptotic behavior depending on the direction.
Such a possibility can be argued only via methods
describing the asymptotic form of the s.p. wave functions appropriately.
\\

\noindent
This work is financially supported in part
as Grant-in-Aid for Scientific Research (C), No.~19540262,
by Japan Society for the Promotion of Science.
Numerical calculations are performed on HITAC SR11000
at Institute of Media and Information Technology, Chiba University,
and at Information Technology Center, University of Tokyo.

\end{document}